\documentclass[aps,prd,12pt,notitlepage,showpacs,nofootinbib,tightenlines]{revtex4-1}
\usepackage{amsmath}
\usepackage{amssymb}
\usepackage{epsfig}
\usepackage{graphicx}
\usepackage{bm}
\usepackage{times}
\usepackage{braket}
\usepackage{color}
\usepackage{slashed}
\usepackage{hyperref}
\definecolor{Red}{rgb}{1.,0.,0.}

\definecolor{Blue}{rgb}{0.,0.,1.}

\definecolor{nicered}{rgb}{0.7,0.1,0.1}
\definecolor{nicegreen}{rgb}{0.1,0.5,0.1}
\hypersetup{colorlinks,citecolor=nicegreen,linkcolor=nicered}

\begin{document}

\newcommand{\beq}{\begin{eqnarray}}
\newcommand{\eeq}{\end{eqnarray}}

\newcommand{\non}{\nonumber\\ }
\newcommand{\ov}{\overline}

\newcommand{\calo}{ {\cal O}}
\newcommand{\calb}{ {\cal B}}
\newcommand{\calr}{ {\cal R}}

\newcommand{\rmt}{ {\rm T}}
\newcommand{\jpsi}{J/\psi}
\newcommand{\etar}{\eta^\prime }
\newcommand{\etap}{\eta^{(\prime)}}
\newcommand{\psl}{ p \hspace{-2.0truemm}/ }
\newcommand{\qsl}{ q \hspace{-2.0truemm}/ }
\newcommand{\epsl}{ \epsilon \hspace{-2.0truemm}/ }
\newcommand{\nsl}{ n \hspace{-2.2truemm}/ }
\newcommand{\vsl}{ v \hspace{-2.2truemm}/ }

\def \cpc{{\bf Chin. Phys. C} }
\def \ctp{{\bf Commun. Theor. Phys} }
\def \epjc{{\bf Eur. Phys. J. C} }
\def \jhep{ {\bf J. High. Energy Phys.} }
\def \jpg{ {\bf J. Phys. G} }
\def \mpla{ {\bf Mod.  Phys. Lett. A} }
\def \npb{ {\bf Nucl. Phys.  B} }
\def \plb{ {\bf Phys.  Lett.  B} }
\def \ppnp{ {\bf Part. Phys $\&$ Nucl. Phys. } }
\def \pr{ {\bf Phys. Rep.} }
\def \prd{ {\bf Phys.Rev.  D} }
\def \prl{ {\bf Phys.Rev.Lett}  }
\def \ptp{ {\bf Prog.Theor.Phys}  }
\def \zpc{ {\bf Z. Phys. C}  }
\def \csb{ {\bf Chin. Sci. Bull.}  }


\title{The $B\to D^{(*)} l\nu_l$ decays in the pQCD approach with the Lattice QCD input }
\author{Ying-Ying Fan$^{1}$, Zhen-Jun Xiao$^{2}$ \footnote{xiaozhenjun@njnu.edu.cn},
Ru-Min Wang$^{1}$, and Bing-Zhong Li$^{1}$ }
\affiliation{$^{1}$ College of Physics and Electronic Engineering, Xinyang Normal University,
 Xinyang 464000,  China\\
$^{2}$Department of Physics and Institute of Theoretical Physics,
Nanjing Normal University, Nanjing 210023,  China}

\begin{abstract}
In this paper, we studied the semileptonic decays $B \to D^{(*)} l^- \bar{\nu}_l$ by using the
``pQCD+Lattice QCD" method.
We made the extrapolation for the six relevant form factors by using
the input values obtained from the pQCD factorization approach in the low $q^2$ region of
$0\leq q^2 \leq m_\tau^2$, and the Lattice QCD input at the end-point $q^2=q^2_{\rm max}$.
We then calculated the ratios $R(D)$ and $R(D^*)$ of the branching ratios
${\cal B}(B \to D^{(*)} l^- \bar{\nu}_l)$, and found numerically that
(1) the ``pQCD+Lattice QCD" predictions for the branching ratios ${\cal B}(B \to D^{(*)} l^-\bar{\nu}_l)$
agree well with the measured values within one standard deviation;
and (2) the ``pQCD+Lattice QCD" predictions  for the ratios $R(D^{(*)})$ are
$ R(D)=0.337^{+0.038}_{-0.037}$ and $ R(D^*)=0.269^{+0.021}_{-0.020}$; they agree with the data
within $2\sigma$ deviation, in other words, one can explain the ``$R(D^{(*)})$-puzzle"
in the framework of the standard model.
\end{abstract}

\maketitle
 keywords: ``$R(D^{(*)})$-puzzle"; $B\to D^{(*)} l\nu_l$ decays; the pQCD approach; Lattice QCD
\section{Introduction}\label{sec:1}

The semileptonic $B$ decays are very important processes in testing the
standard model (SM) and in searching for the new physics (NP).
The $B \to D^{(*)}l \bar{\nu}_l$ semileptonic decays have been measured by
the BaBar in 2012 \cite{babar2008,belle2007,belle2010,prl109-101802,prd88-072012},
and by the Belle and LHCb Collaborations this year with different methods~
\cite{belle2015a,belle2015b,lhcb-2015}.
For the ratios of the branching ratios: $ R(D^{(*)}) \equiv  {\cal B}(B \to D^{(*)} \tau \bar{\nu}_\tau)
/{\cal B}(B \to D^{(*)} l \bar{\nu}_l)$ with $l = (e, \mu)$, the individual measurements
\cite{prl109-101802,prd88-072012,belle2015a,belle2015b,lhcb-2015} and the world averages
\cite{hfag-2015} are the following:
\beq
{\cal R}(D)^{\rm exp} &=& \left\{ \begin{array}{ll}
0.440 \pm 0.072, \; & {\rm BaBar}\; [4,5], \\
0.375 ^{+0.064}_{-0.063}\pm 0.026, \; & {\rm Belle}\;[6,7], \\
0.391 \pm 0.041 \pm 0.028, \; & {\rm HFAG \ \ average}\; [9] , \\ \end{array} \right. \label{eq:exp01}\\
{\cal R}(D^*)^{\rm exp} &=& \left\{ \begin{array}{ll}
0.332 \pm 0.024\pm 0.018, \; & {\rm BaBar}\;[4,5], \\
0.293 ^{+0.039}_{-0.037}\pm 0.015, \; & {\rm Belle}\; [6,7], \\
0.336\pm 0.027 \pm  0.030, \; & {\rm LHCb}\; [8], \\
0.322 \pm 0.018\pm 0.012, \; & {\rm HFAG \ \ average}\; [9] . \\ \end{array} \right. \label{eq:exp02}
\eeq

On the theory side, the standard method to study the semileptonic $B\to D^{(*)} l
\bar{\nu}$ decays is the heavy-quark-effective-theory (HQET). The SM predictions based on
the HQET as given in Ref.~~\cite{prd85-094025} are the following:
\beq
{\cal R}(D) &=& 0.296\pm 0.016\; , \quad {\cal R}(D^*) = 0.252\pm 0.003.
\label{eq:sm01}
\eeq
Two new lattice calculations for the ratio $R(D)$ \cite{l2015-1,l2015-2} are of the form
\beq
{\cal R}(D) &=& 0.299\pm 0.011\;, \quad {\cal R}(D) = 0.300\pm 0.008\;.
\label{eq:lattice}
\eeq

By comparing the measured values of $R(D^{(*)})$ in Eqs.~(\ref{eq:exp01}) and (\ref{eq:exp02})
with those SM expectations in Eqs.~(\ref{eq:sm01}) and (\ref{eq:lattice}), one can see the following points:
\begin{enumerate}
\item[(1)]
For BaBar data only, the measured values of $R(D^{(*)})$ disagree with the SM predictions
in Eq.~(3) by about $3.4\sigma$ \cite{hfag-2015}, this is the so-called ``$R(D^{(*)})$-puzzle".

\item[(2)]
For the ratio $R(D)$ and $R(D^*)$, the deviation of $1.8\sigma$ of the Belle data from
the HQET predictions of $R(D^{(*)})$ \cite{belle2015a,belle2015b}  is by  itself
not significant, but the Belle results do increase the tension
of the world average values with the HQET predictions slightly.
A combination for both BaBar and Belle measurements
made in Refs.~\cite{belle2015a,belle2015b} showed  that the deviation can reach $4.0 \sigma$.

\item[(3)]
The LHCb reported very recently its first measurement for the ratio $R(D^*)$ \cite{lhcb-2015}, the job
considered unfeasible at hadron colliders before. The LHCb measurement of $R(D^*)=0.336\pm 0.027 \pm 0.030$
itself agrees with SM (HQET) at $2.1\sigma$ level, while the combined HFAG averages of $R(D)$ and $R(D^*)$ as
given in Eqs.(1) and (2) do deviate from the SM predictions in Eqs.~(3) and (4) at $3.9\sigma$ level
\cite{hfag-2015}.

\end{enumerate}

How to understand the above ``$R(D^{(*)})$-puzzle" ? which has invoked intensive studies
in the framework of the SM and/or  various NP models, for example, in
Refs.~\cite{prl109-071802,prd85-114502,mpla27-1250183,prl109-161801,
prd86-054014,prd86-034027,prd86-114037,plb742-370,plb716-208,prd87-054002,prd87-014014,
jhep2013-01054,prd87-074010,csb59-125,csb59-3787,ligeti15a,ligeti13a}.
As is well-known, the theoretical predictions for the ratios $R(D)$ and $R(D^*)$ have
a direct strong dependence on the values and the shapes of the $B \to D^{(*)}$ transition form factors
$F_{(+,0)}(q^2)$, $V(q^2)$ and $A_{0,1,2}(q^2)$.

In Ref.~\cite{csb59-125}, we calculated the ratios $R(D^{(*)})$ by employing the pQCD factorization approach,
where the relevant hadronic form factors are evaluated in the low $q^2$ region by employing the pQCD factorization
approach and then extrapolated to the higher $q^2$ region by using the pole model
parametrization ~\cite{prd79-054012}.
The resulted pQCD predictions are ${\cal R}(D) = 0.430^{+0.021}_{-0.026}$, ${\cal R}(D^*) = 0.301 \pm 0.013$
\cite{csb59-125}. These pQCD predictions agree very well with the BaBar results in
Eqs.~(\ref{eq:exp01}) and (\ref{eq:exp02}), but we also know that the pQCD  predictions for $B\to D^{(*)}$
transition form factors are reliable only in the low $q^2$ region
\cite{csb59-125,prd67-054028,ppnp51-85,xiaobc,zou2b},
and the extrapolation from the pQCD  predictions at low $q^2$ region to the high $q^2$ region by using the pole
model parametrization  may have large theoretical uncertainties.
We need something new to improve the reliability of the extrapolation method,
the Lattice QCD input of $F(q^2 =q^2_{\rm max})$ is what we need to determine
reliably the size and the shape of all the six form factors $F(q^2)$ in the whole
range of $0 < q^2 \leq q^2_{\rm max}$.

In Ref.~\cite{prl109-071802}, for example, the authors calculated the ratio
$R(D)$ by using the hadronic form factors computed in lattice QCD
in the high $q^2$ region and found $R(D)=0.316\pm 0.012({\rm stat.}) \pm 0.007({\rm syst.})$,
which reduced slightly the tension between the SM prediction in Eq.~(\ref{eq:sm01}) and
the BaBar result in Eq.~(\ref{eq:exp01}).

In this paper, we firstly calculate the form factors $F(q^2)$ in the lower $q^2\leq
m^2_{\tau}$ region by using the pQCD factorization approach,
and then take the $F(q^2 = q^2_{\rm max})$ evaluated in the
Lattice QCD as additional input to improve the reliability of the pole model
extrapolation, and finally to calculate the branching ratios
${\cal B}(B \to D^{(*)} l^- \bar{\nu}_l)$ and the ratios $R(D^{(*)})$.


\section{Form factors and their extrapolations}\label{sec:2}

In the $B$ meson rest frame, we define the $B$ meson momentum $P_1$,
the $D^{(*)}$ momentum $P_2$, and the polarization
vectors $\epsilon$ of the $D^{*}$ in the light-cone coordinates in the same way as in
Refs.~\cite{prd67-054028,csb59-125}.
For the $B$ meson wave function, we use the same one as being used for example in
Ref.~\cite{csb59-125}:
\beq
\Phi_{B}=\frac{\rm i}{\sqrt{2N_{\rm c}}} (\psl_{B} +m_{B}) \gamma_5 \phi_{B} (x,b),
\label{eq:bmeson}
\eeq
with the distribution amplitude $\phi_B$ in the form of
\beq
\phi_{B}(x,b)&=& N_{B} x^2(1-x)^2\mathrm{\exp} \left
 [ -\frac{M_{B}^2\; x^2}{2 \omega_{B}^2} -\frac{1}{2} (\omega_{B} b)^2\right],
\label{eq:phib}
\eeq
where the normalization factor $N_{B}$ depends on the values
of the parameter $\omega_{B}$ and decay constant $f_{B}$ and is defined
through the normalization relation:
$ \int_0^1{\rm d}x\; \phi_{B}(x,b=0)=f_{B}/(2\sqrt{6})$.
We set the shape parameter $\omega_B =(0.40\pm 0.04)$~GeV, in order to estimate
the uncertainties of theoretical predictions.

For the $D$ and $D^*$ meson, we use the same wave functions as being used in
Refs.~\cite{prd78-014018,csb59-125}
\beq
\Phi_{D}(p,x)&=&\frac{\rm i}{\sqrt{6}}\gamma_5 (\psl_{D}+ m_{D} )\phi_{D}(x), \label{eq:wfd} \\
\Phi_{D^*}(p,x) &=& \frac{- \rm i}{\sqrt{6}} \left
 [  \epsl_{\rm L}(\psl_{D^*} +m_{D^*})\phi^{\rm L}_{D^*}(x)
 + \epsl_{\rm T}(\psl_{D^*} + m_{D^*})\phi^{\rm T}_{D^*}(x)\right ],
\label{eq:wfdstar}
 \eeq
with the distribution amplitudes
\beq
\phi_{ D^{(*)} }(x)=\frac{ f_{D^{(*)} }}{2\sqrt{6}} 6x(1-x)
\left[ 1+C_{D^{(*)} }(1-2x)\right]\exp
\left[-\frac{\omega^2 b^2 }{2}\right], \label{eq:phid}
\eeq
and with the choices of $f^{\rm L}_{D^*}=f^{\rm T}_{D^*}=f_{D^*}$,~$\phi^{\rm L}_{D^*}=\phi^{\rm T}_{D^*}=\phi_{D^*}$,
and $C_{D}=C_{D^*}=0.5,~ \omega=0.1$ GeV as in
Refs.~\cite{prd78-014018,csb59-125}.


In the framework of the SM, the differential decay widths
of the semileptonic $B \to D l\bar{\nu}_l$ decays can be written as~\cite{plb716-208}
\beq
\frac{{\rm d}\mathnormal{\Gamma}(B \to D l\bar{\nu}_l)}{{\rm d}q^2}&=&\frac{G_{\rm F}^2|V_{cb}|^2}{192 \pi^3  m_{B}^3}
\left ( 1-\frac{m_l^2}{q^2} \right)^2\frac{
\lambda^{1/2}(q^2)}{2q^2}\non && \cdot \Bigl \{  3 m_l^2\left
(m_{B}^2-m_{D}^2 \right )^2 |F_0(q^2)|^2 \non && + \left
(m_l^2+2q^2 \right )\lambda(q^2)|F_+(q^2)|^2 \Bigr \},
\label{eq:dg1}
\eeq
where $m_l$ is the mass of the relevant leptons, $\lambda(q^2) = (m_B^2+m_{D}^2-q^2)^2 - 4 m_B^2 m_{D}^2$
is the phase space factor, $F_{(+,0)}(q^2)$ are the form factors.

For $B \to D^* l\bar{\nu}_l$ decays, the corresponding differential decay
widths can be written as~\cite{prd79-054012}
\beq
\frac{{\rm d}\mathnormal{\Gamma}_{\rm L}(\bar{B}^0 \to D^{*} l
\bar{\nu}_l)}{{\rm d}q^2}&=& \frac{G_{\rm F}^2|V_{cb}|^2}{192 \pi^3  m_{B}^3}
\left ( 1-\frac{m_l^2}{q^2}\right )^2
\frac{\lambda^{1/2}(q^2)}{2q^2}\non
&&\cdot \Bigg\{3m^2_l\lambda(q^2)A^2_0(q^2)+\frac{m^2_l+2q^2}{4m_{D^*}^2}\non
&&
\cdot \left  [(m^2_{B}-m_{D^*}^2-q^2)(m_{B}+m_{D^*})A_1(q^2)
\right. \non
&&  \left.-\frac{\lambda(q^2)}{m_{B}+m_{D^*}}A_2(q^2) \right  ]^2 \Bigg\},
\label{eq:dfds1}
\eeq
\beq
\frac{{\rm d}\mathnormal{\Gamma}_\pm(\bar{B}^0 \to D^{*} l\bar\nu_l)}{{\rm d}q^2}&=& \frac{G_{\rm F}^2|V_{cb}|^2}{192 \pi^3
m_{B}^3}\left ( 1-\frac{m_l^2}{q^2}\right )^2
\frac{\lambda^{3/2}(q^2)}{2}\non & & \times  \left \{
(m^2_l+2q^2)\left[\frac{V(q^2)}{m_{B}+m_{D^*}}\right. \right.\non & & \left. \left.\mp
\frac{(m_{B}+m_{D^*})A_1(q^2)}{\sqrt{\lambda(q^2)}}\right]^2\right\},
\label{eq:dfds2}
 \eeq
where $\lambda(q^2) = (m_{B}^2+m_{D^*}^2-q^2)^2 - 4 m_{B}^2 m_{D^*}^2$ is the
phase space factor, $V(q^2)$ and $A_{0,1,2}(q^2)$ are the relevant form factors.
The total differential decay widths can be defined as
 \beq
\frac{{\rm d}\mathnormal{\Gamma}}{{\rm d}q^2}=\frac{{\rm d}\mathnormal{\Gamma}_{\rm L}}{{\rm d}q^2} + \frac{{\rm d}\mathnormal{\Gamma}_+}{{\rm d}q^2} +\frac{{\rm d}\mathnormal{\Gamma}_-}{{\rm d}q^2}. \label{eq:dfdst}
\eeq

The traditional methods to evaluate the form factors for $B\to D^{(*)}$ transitions
are the QCD sum rule  and the Lattice QCD. Each method has its own advantage region:
the QCD sum rule are reliable in the low $q^2$ region, but the Lattice QCD in
the high $q^2$ region of $q^2\approx q_{\rm max}^2$.
In Refs.~\cite{prd67-054028,prd78-014018}, the authors examined the applicability of the pQCD approach to
$B \to D^{(*)}$ transitions, and have shown that the pQCD approach based on the $k_{\rm{T}}$ factorization theorem to the
$B \to D^{(*)}$ transitions is reliable in the lower $q^2$ region.
In Ref.~\cite{csb59-125,prd89-014030}, we evaluated the form factors for $B/B_s \to (D^{(*)},D_{s}^{(*)})$
transitions in the lower $q^2$ region and obtained the pQCD predictions for the ratios
$R(D^{(*)})$ and $R(D_s^{(*)})$.

In order to improve the reliability of the extrapolation of the form factors from the low to high $q^2$ region, we here take the Lattice QCD predictions for all six relevant form factors  at the end-point $q^2=q^2_{\rm max}$ as the additional
input in the fitting process.

The parametrization of the form factors are rather different in different theories or models.
For $B \to D$ transition, the form factors $h_{+,-}(w)$ ( or ${\cal G}(w)$ ) in the Lattice QCD
has simple relations with the form factors $F_{+,0}(q^2)$ in the QCD sum rule and/or the pQCD
approach \cite{prd85-114502,plb716-208,csb59-3787}
\beq
F_+(q^2)&=& \frac{1}{2\sqrt{r}} \left [ (1+r) h_+(w) - (1-r)h_-(w) \right] = \frac{1+r}{2\sqrt{r}} \;  {\cal G}(w) ,
\label{eq:rl03}\\
F_0(q^2) &=&   \sqrt{r} \left [ \frac{1+w}{1+r}h_+(w) -\frac{w-1}{1-r} h_-(w) \right]
= \frac{\sqrt{ m_B m_D }}{m_B - m_D} {\cal G}(w)\; \mathnormal{\Delta}(w),
\label{eq:rl04}
\eeq
where $r=m_D/m_B$, $w= (m_B^2+m_{D}^2 -q^2)/(2m_B m_{D})$ with $q^2=(p_B-p_D)^2$.
The scalar density $\mathnormal{\Delta}(w)$ in Eq.~(\ref{eq:rl04})
is approximated by a constant value $\mathnormal{\Delta}(w) = 0.46 \pm 0.02$ \cite{prd85-094025,jhep2013-01054}.

For the $\bar B \to D^*$ transition, similarly, the form factors
$V(q^2)$ and $A_{0,1,2}(q^2)$ are related to the Lattice QCD ones  $h_{A_i}(w)$ through the following
relations~\cite{prd85-094025,prd89-114504}
\beq
V(q^2)&=&\frac{1+r}{2\sqrt{r}}h_V(w),\non
A_0(q^2)&=&\frac{1}{2\sqrt{r}}\left [ (1+w)h_{A_1}(w)-(1-w r)
h_{A_2}(w)+(r-w)h_{A_3}(w) \right ],\non
A_1(q^2)&=&\frac{\sqrt{r}}{1+r}(1+w)h_{A_1}(w),\non
A_2(q^2)&=&\frac{1+r}{2\sqrt{r}}(r h_{A_2}(w)+h_{A_3}(w)),
\label{eq:V-A2}
\eeq
where $r=m_{D^*}/m_B$, $w= (m_B^2+m_{D^*}^2 -q^2)/(2m_B m_{D^*})$ with $q^2=(p_B-p_{D^*})^2$,
the explicit expressions of  $h_V(w)$ and $h_{A_i}(w)$ can be found easily in Refs.~\cite{prd85-094025,prd89-114504}.

At the zero-recoil limit: $w\to 1$ and $q^2=q^2_{\rm max}$, we found simple relations $h_V(1)=h_{A_1}(1)=h_{A_3}(1)$
and $h_{A_2}(1)=0$, and  therefore the six form factors at the end-point $q^2=q^2_{\rm max}$
, say $F_{+,0}(q^2_{\rm max}), V(q^2_{\rm max})$ and $A_{0,1,2}(q^2_{\rm max})$, depend on two Lattice QCD
parameters ${\cal G}(1)$ and $h_{A_1}(1)$ only.
By using the formulae as given in Eqs.~(\ref{eq:rl03})$-$(\ref{eq:V-A2})
and the following two Lattice QCD inputs~\cite{prd85-114502,prd89-114504}:
\beq
{\cal G}(1)=1.058\pm 0.009, \quad h_{A_1}(1)={\cal F}(1)= 0.906\pm 0.004\pm 0.012,
\label{eq:g1-ha1}
\eeq
one can find the values of the relevant form factors at the end-point $q^2=q_{\rm max}^2$,
\beq
F_0(q^2_{\rm max})&=& 0.92 \pm 0.02 ,\quad F_+(q^2_{\rm max})= 1.21 \pm 0.02, \quad  V(q^2_{\rm max})= 1.01 \pm 0.02,\non
A_0(q^2_{\rm max})&=& 1.01 \pm 0.02,\quad A_1(q^2_{\rm max})= 0.81 \pm 0.02,  \quad A_2(q^2_{\rm max})= 1.01 \pm 0.02,
\label{eq:lqcds}
\eeq
where $q_{\rm max}^2=(m_B-m_D)^2=11.63$ GeV$^2$ for $B\to D$ transition, $q_{\rm max}^2=(m_B-m_{D^*} )^2=10.69$ Gev$^2$
for $B\to D^*$ transition. The uncertainty of the form factors induced by the errors of
${\cal G}(1)$ and $h_{A_1}(1)$ as given in Eq.~(\ref{eq:g1-ha1}) is around $0.01$ only.
We here set conservatively the common error of $\pm 0.02$  for all six form factors in Eq.~(\ref{eq:lqcds})
by taking into account approximately the small variations of the central values of ${\cal G}(1)$ and $h_{A_1}(1)$
in recent years~\cite{prd85-114502,prd89-114504,l2015-1}.

\begin{figure}[thb]
\begin{center}
\vspace{-0.5cm}
\centerline{\epsfxsize=6cm\epsffile{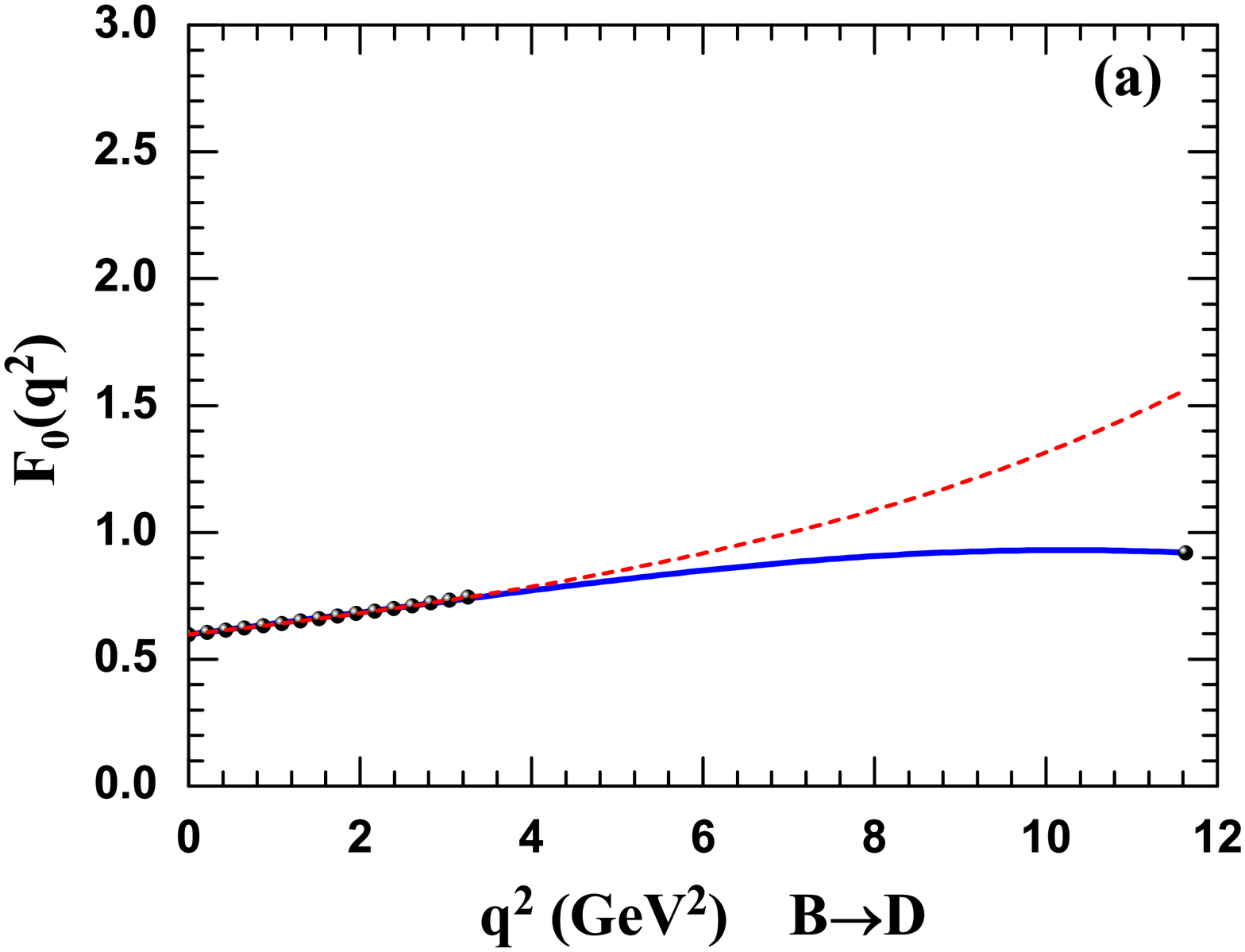}
\hspace{-0.8cm} \epsfxsize=6cm\epsffile{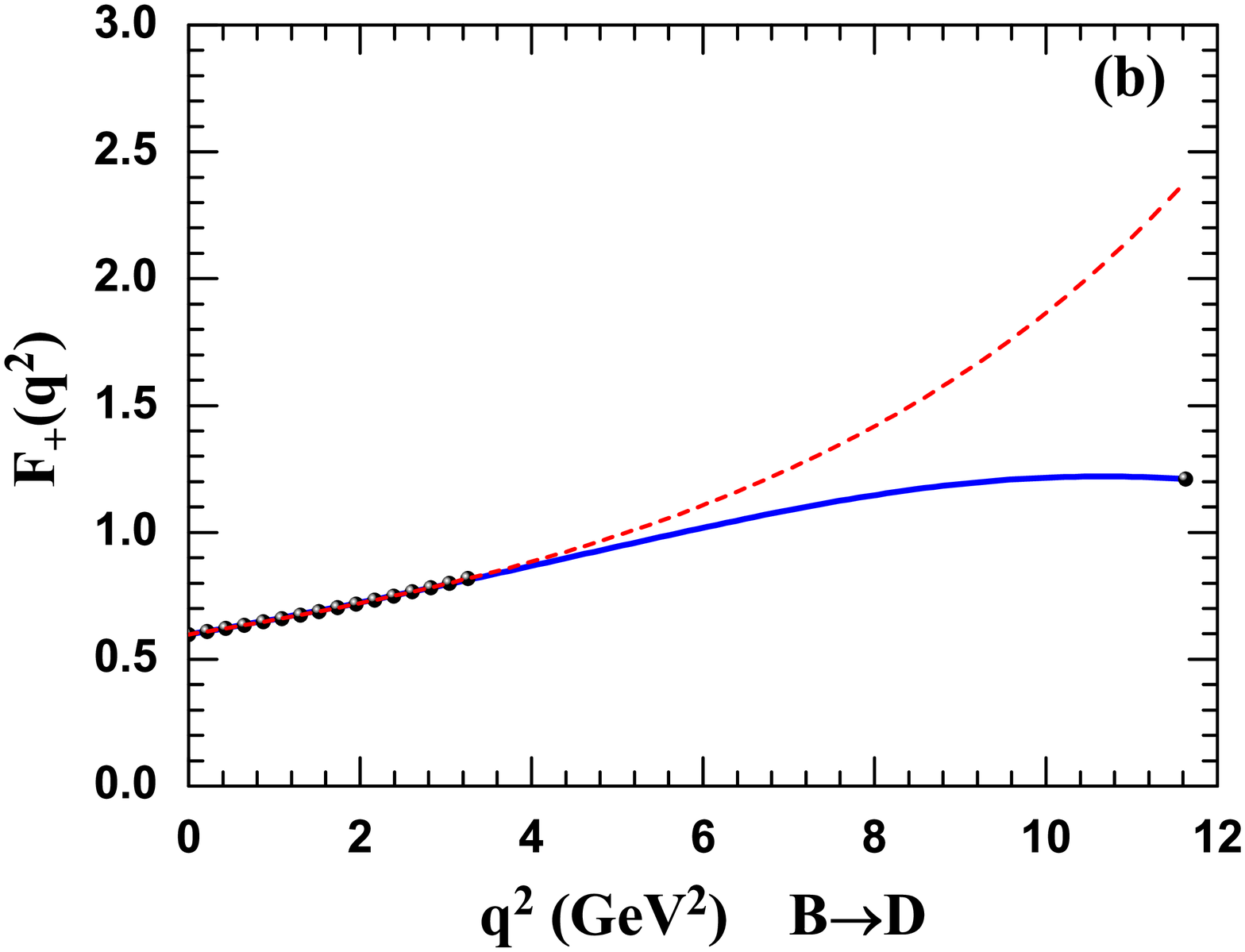}
\hspace{-0.8cm} \epsfxsize=6cm\epsffile{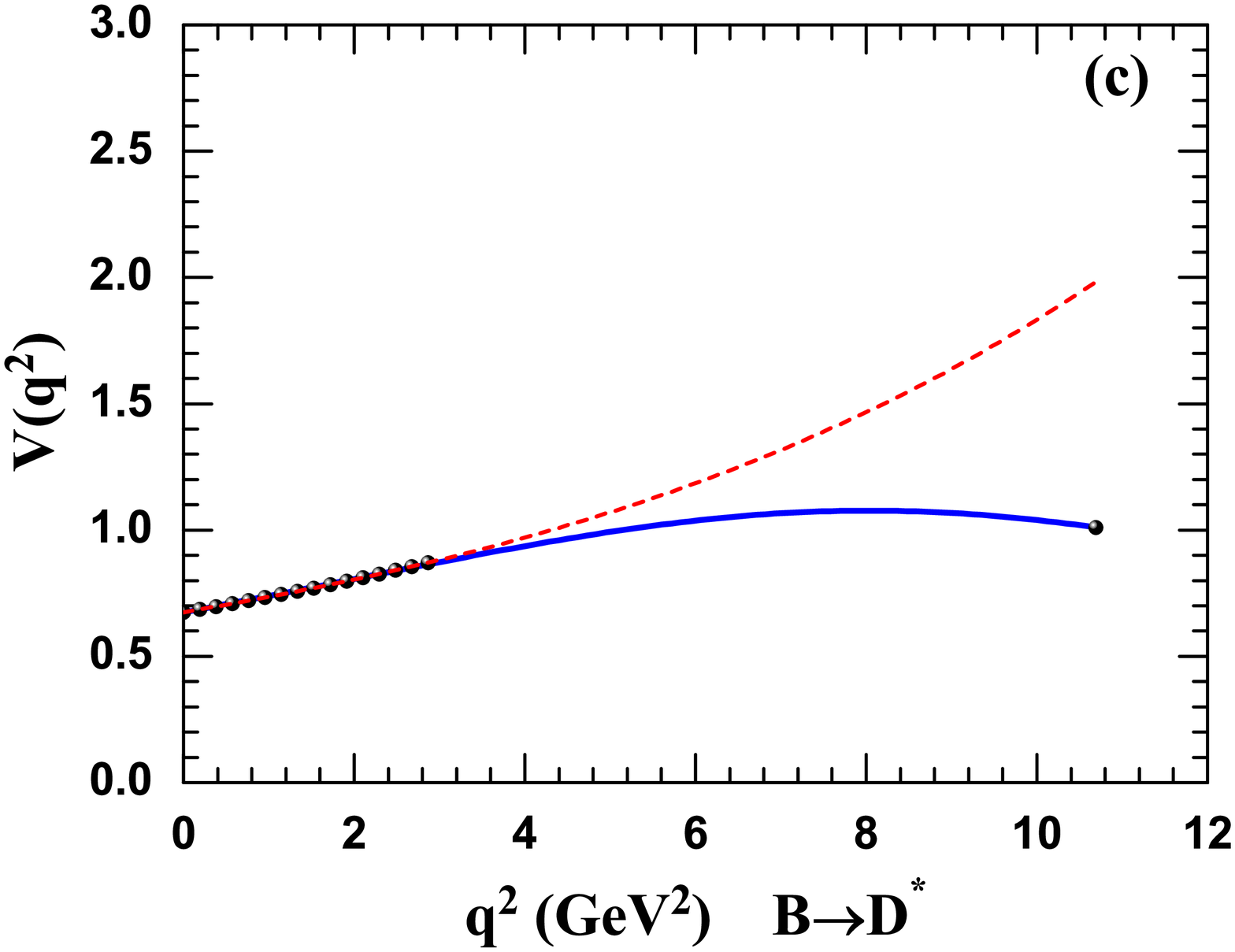}}
\vspace{0.3cm}
\centerline{\epsfxsize=6cm\epsffile{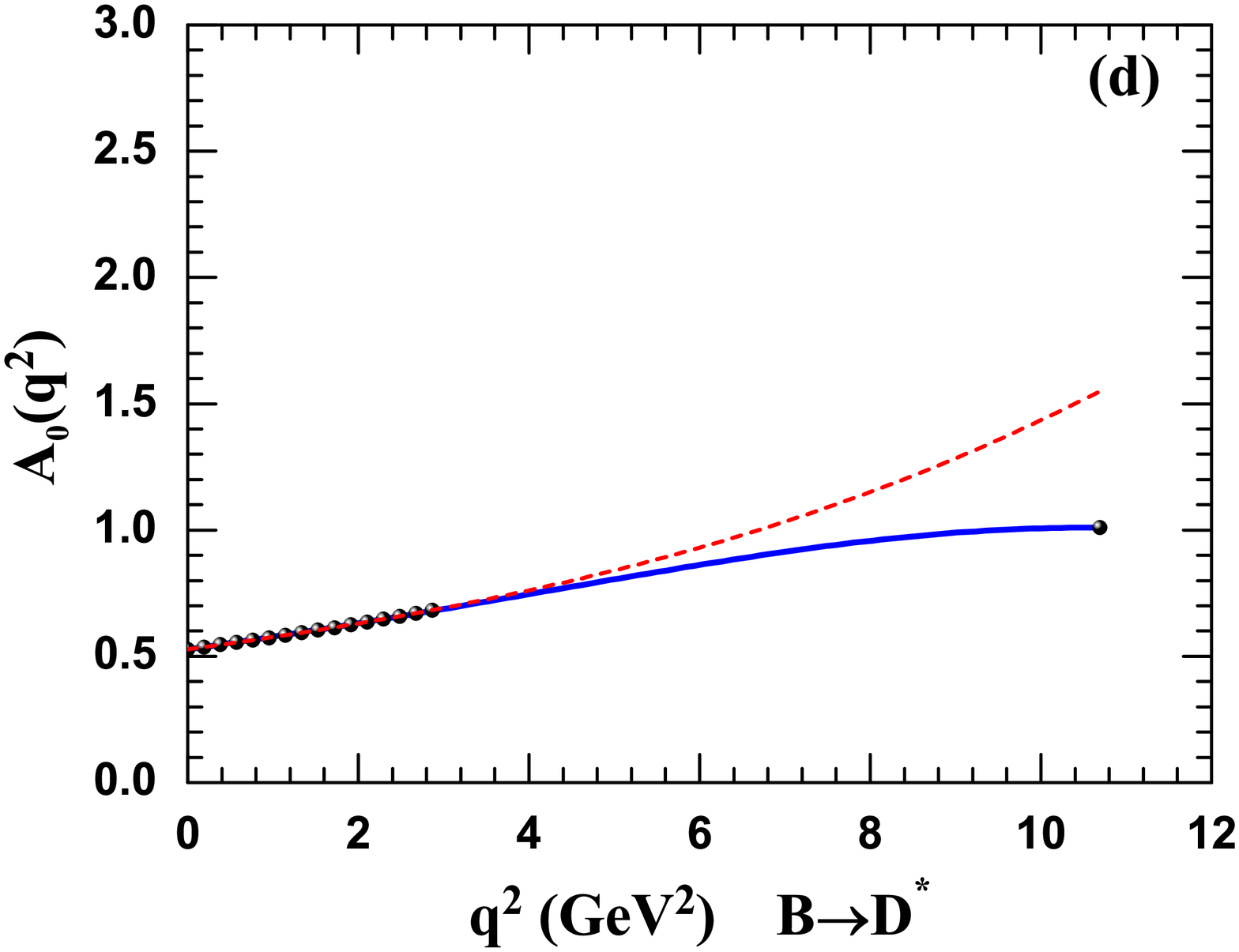}
\hspace{-0.8cm}\epsfxsize=6cm\epsffile{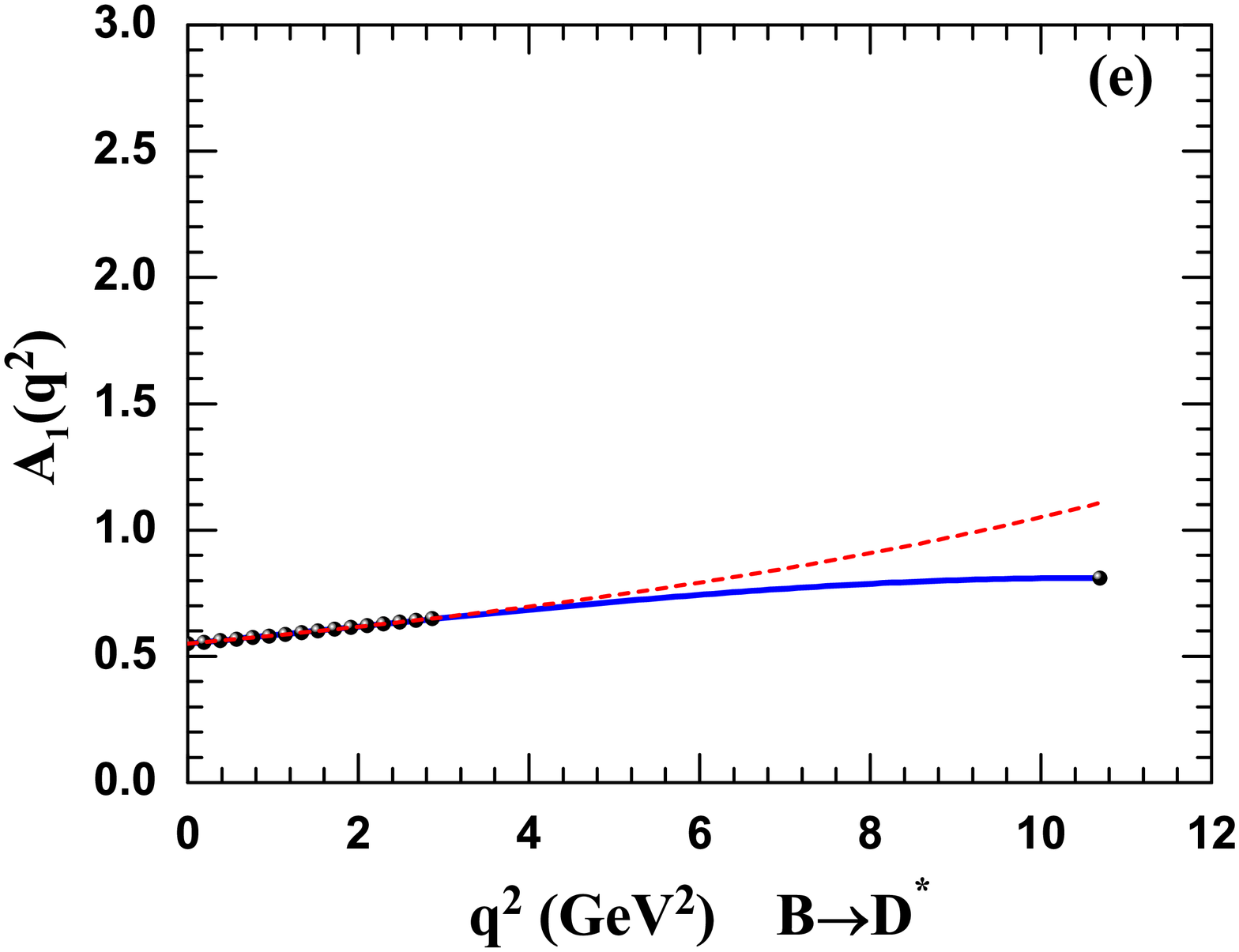}
\hspace{-0.6cm}\epsfxsize=6cm\epsffile{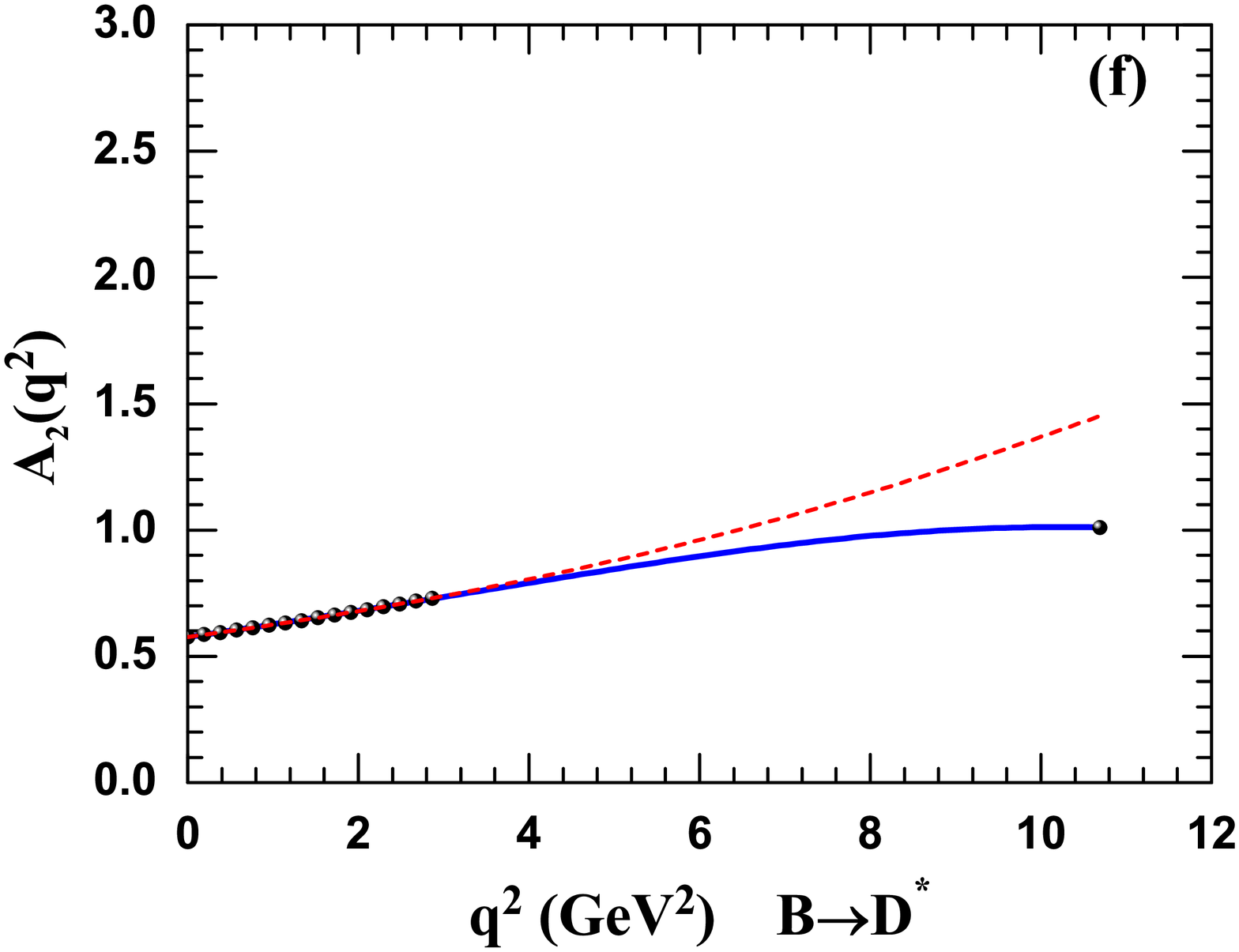}  }
\caption{(Color online) The theoretical predictions for the $q^2$-dependence of the six form factors
in the pQCD approach (the dashed curves), and the "pQCD+Lattice QCD" method (solid curves). }
\label{fig:fig1} \end{center}
\end{figure}

Now we make the extrapolation from the low $q^2$ to the higher $q^2$ region by
using the pole model parametrization~\cite{prd79-054012}:
\beq
F(q^2)=\frac{F(0)}{1-a (q^2/m_B^2) + b\left( q^2/m_B^2\right)^2},
\label{eq:fq2}
\eeq
where $F(q^2)$ stands for all the six relevant form factors $F_{0,+}(q^2)$, $V(q^2)$
and $A_{0,1,2}(q^2)$. The parameters $a$ and $b$ in the above equation are determined
by the fitting to the values at the seventeen points: the first sixteen points in the lower $q^2$
region of $q^2 \leq m_\tau^2$ calculated by using the pQCD factorization approach,
and the last one at the end-point $q^2=q_{\rm max}^2$ from the Lattice QCD approach in Eq.~(\ref{eq:lqcds}),
as illustrated explicitly in Fig.~\ref{fig:fig1}.

In Fig.~\ref{fig:fig1}, the dashed curves show the theoretical predictions for the $q^2$-dependence of
the six relevant form factors in the pQCD approach only \cite{csb59-125},
while the solid ones denote
the six form factors evaluated by using the  "pQCD + Lattice QCD" method.
One can see clearly that: both the values and the shapes of all six form factors changed
explicitly, in the higher $q^2$ region of $7 \leq q^2 \leq q^2_{\rm max}$,
because of the inclusion of the Lattice QCD input as listed in Eq.~(\ref{eq:lqcds})
at the end-point $q^2=q^2_{\rm max}$.

\section{Numerical results}\label{sec:3}

In the numerical calculations we use the following input parameters
(here masses and decay constants in units of GeV)~\cite{pdg2014}:
\beq
m_{D^0}&=&1.865, \quad m_{D^+}=1.870, \quad m_{D^{0*}}=2.007,\quad m_{D^{*+}}=2.010,\non
m_B&=&5.28, \quad \quad m_{\tau}= 1.777,\quad m_{c}=1.35 \pm 0.03, \quad  f_B=0.21\pm0.02, \quad f_{D}=0.223,  \non
|V_{cb}|&=&(39.5 \pm 0.8)\times 10^{-3}, \quad {\mathnormal{\Lambda}}^{(f=4)}_{\rm \bar{MS}} = 0.287,\quad
\tau_{B^\pm}= 1.638\; {\rm ps},\quad \tau_{B^0}=1.519\; {\rm ps},\label{eq:inputs}
\eeq
while $f_{D^*}= f_{D} \sqrt{m_{D}/m_{D^*}}, \quad f_{D_{s}^*}= f_{D_{s}} \sqrt{m_{D_{s}}/m_{D_{s}^*}}$.

\begin{figure}[tb]
\begin{center}
\vspace{-0.5cm}
\centerline{\epsfxsize=9cm\epsffile{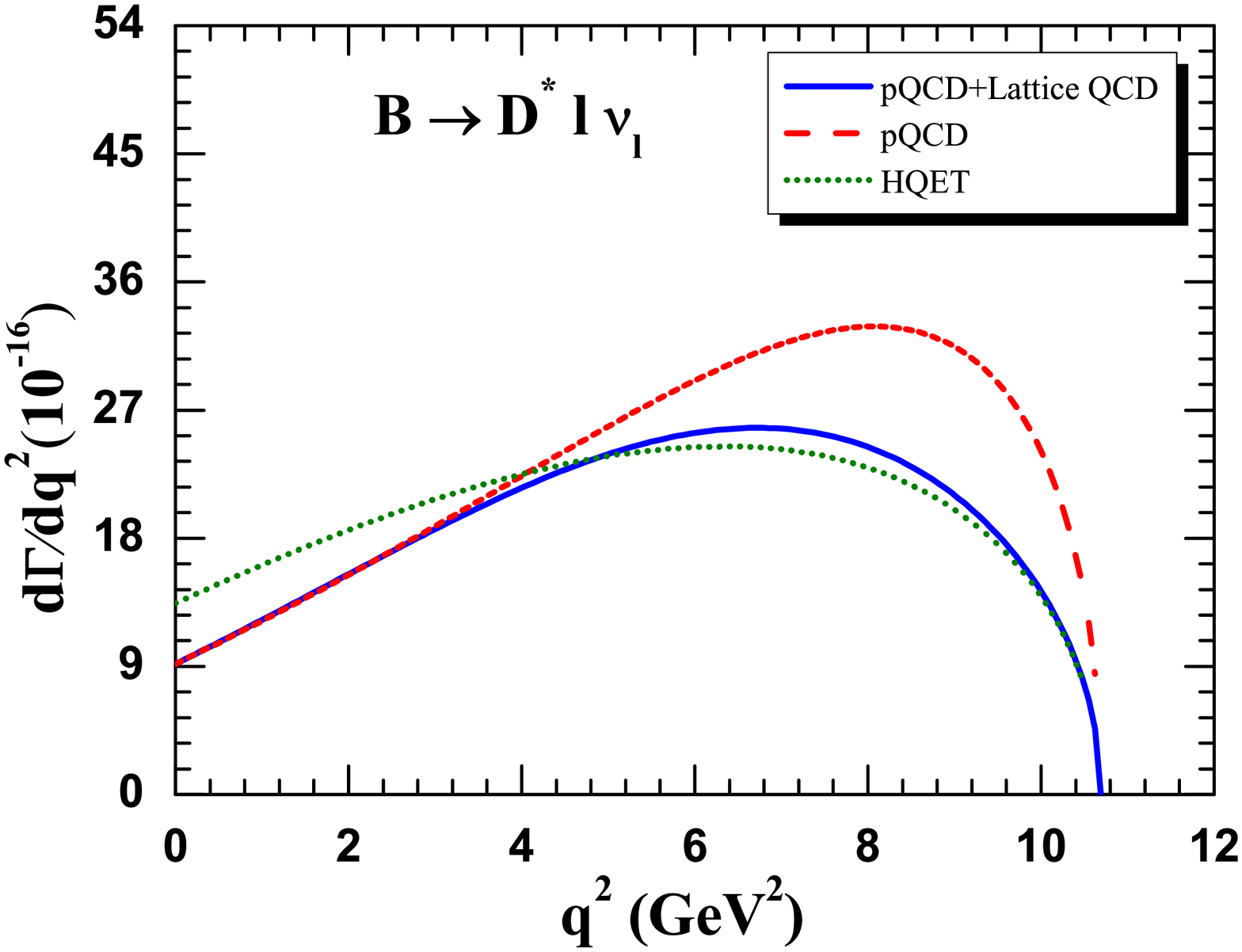}
\hspace{-1.2cm} \epsfxsize=9cm\epsffile{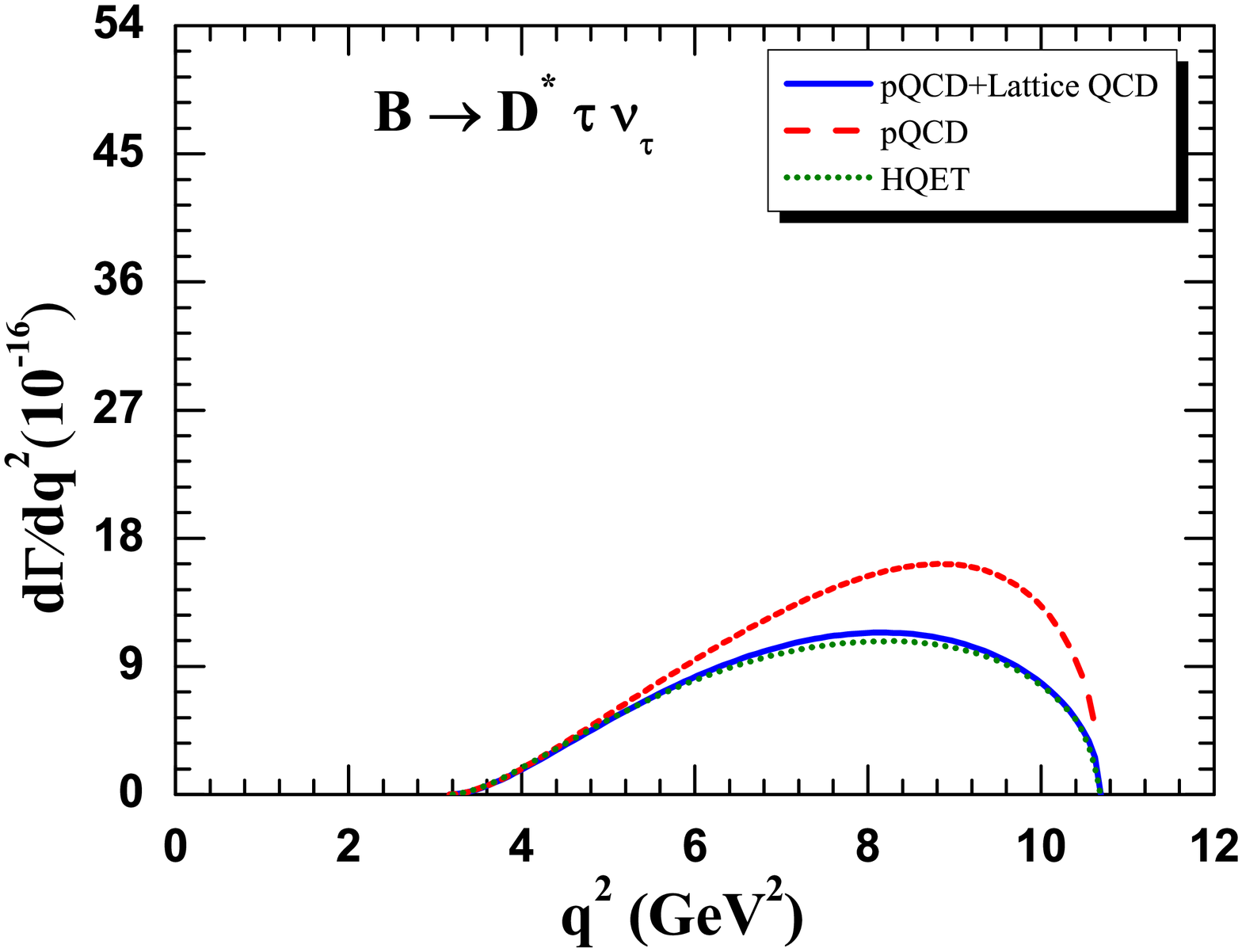}}
\vspace{0.3cm}
\caption{(Color online) The theoretical predictions for the $q^2$-dependence of ${\rm d}{\mathnormal{\Gamma}}/{\rm d}q^2$ for
$B \to D^* l^-\bar{\nu}_l$ $l=(e,\mu)$ and $B \to D^* \tau^-\bar{\nu}_\tau$ decays
in the pQCD approach, the ``pQCD + Lattice-QCD" method, and
the traditional HQET method. Here $q^2_{\rm max} = 10.69$ GeV$^2$}
\label{fig:fig2}
\end{center}
\end{figure}

In Fig.~\ref{fig:fig2}, furthermore, we show the $q^2$-dependence of the
theoretical predictions for the differential decay rates ${\rm d}{\mathnormal{\Gamma}}/{\rm d}q^2$ for
$B\to D^{*} l \bar{\nu}_l$ with $l=(e^-,\mu^-)$ and $B\to D^{*} \tau \bar{\nu}_\tau$ decays
calculated by using the pQCD approach only (the dashed curves ) \cite{csb59-125},
the traditional HQET method \cite{prd85-094025} (the short-dots curve) and
the new ``pQCD + Lattice QCD" method, respectively.

From the curves in Figs.~\ref{fig:fig1} and \ref{fig:fig2} one can see that
\begin{enumerate}
\item[(1)]
For all the six form factors,  the $q^2$-dependence of the ``pQCD+Lattice QCD" predictions are
similar with those pQCD ones in the lower $q^2$ region, but become rather different in both
the shape (flat) and the magnitude (smaller) from the previous ones in the large $q^2$ region
due to the strong suppression effect of the Lattice QCD input at the end-point $q^2=q^2_{\rm max}$
in the extrapolation.

\item[(2)]
For the considered differential decay rates, the  ``pQCD+Lattice QCD" predictions
become much smaller than the previous pure pQCD predictions in the large $q^2$ region, but
approach to the HQET ones in the large $q^2$ region.
\end{enumerate}

In Table \ref{tab:br31} we list the theoretical predictions for the branching ratios of the
considered eight semileptonic $B$ decays, where the theoretical errors from different
sources (say $\omega_B=0.40\pm 0.04$ GeV, $f_B=0.21\pm 0.02$ GeV,
$|V_{cb}|=(39.5\pm 0.8) \times 10^{-3}$ and $m_c=1.35\pm 0.03$ GeV) have been added in quadrature.
As a comparison, we also show the theoretical predictions obtained by using the traditional
HQET method \cite{prd85-094025} or the pQCD factorization approach only \cite{csb59-125}.
The world averages as given in PDG 2014 ~\cite{pdg2014} are also listed in last column.
One can see from the numerical results as shown in Table~\ref{tab:br31} that
\begin{enumerate}
\item[(1)]
For all eight decays, the theoretical prediction from the pQCD approach \cite{csb59-125} and
the ``pQCD+Lattice QCD" method agree well with each other within one standard deviation.
They are also well consistent with the world averages as listed in last column \cite{pdg2014}.

\item[(2)]
For the branching ratios of the considered semileptonic decays, there is no obvious deviations
between the theoretical predictions and the measured ones, since the corresponding
theoretical errors are still large, around $\sim 30~\%$ in size and can not be reduced soon.

\end{enumerate}

\begin{table}[tb]
\begin{center}
\caption{ The theoretical predictions for the branching ratios (in units of $10^{-2}$) of
the $B \to D^{(*)} l \nu_l$ and $B \to D^{(*)} \tau \nu_\tau$ decays.
The previous theoretical predictions in Refs.~\cite{prd85-094025,csb59-125}
and the world averages in Ref.~\cite{pdg2014} are also listed as a comparison.}
\label{tab:br31}\vspace{0.2cm}
\begin{tabular}{l l c l l} \hline
 Channel \qquad& pQCD \cite{csb59-125}\quad \qquad& pQCD+Lattice QCD \qquad   & HQET \cite{prd85-094025}\qquad&
 PDG \cite{pdg2014} \quad \\
\hline  
$\bar{B}^0 \to D^+ \tau^- \bar{\nu}_\tau$ & $0.87^{+0.34}_{-0.28}$& $0.78^{+0.23}_{-0.20}$ &$0.57-0.69 $             & $1.03\pm 0.22$\\
$\bar{B}^0 \to D^+ l^- \bar{\nu}_l$       & $2.03^{+0.92}_{-0.70}$& $2.31^{+1.05}_{-0.77}$ &$2.13 ^{+0.19}_{-0.18}$  & $2.19\pm 0.12$\\
$B^- \to D^0 \tau^- \bar{\nu}_\tau$       & $0.95^{+0.37}_{-0.31}$& $0.84^{+0.25}_{-0.21}$ &$0.61-0.75 $             & $0.77\pm 0.25$\\
$B^- \to D^0 l^- \bar{\nu}_l$             & $2.19^{+0.99}_{-0.76}$& $2.48^{+1.12}_{-0.83}$ &$2.30 \pm 0.20$          & $2.27\pm 0.11$\\
$\bar{B}^0 \to D^{*+}\tau^-\bar{\nu}_\tau$& $1.36^{+0.38}_{-0.37}$  & $1.29^{+0.25}_{-0.24}$ &$1.21-1.35$    & $1.84\pm 0.22$\\
$\bar{B}^0\to D^{*+} l^- \bar{\nu}_l$     & $4.52^{+1.44}_{-1.31}$  & $4.79^{+1.37}_{-1.18}$ &$4.94\pm 0.15$ & $4.93\pm 0.11$\\
$B^-\to D^{*0} \tau^- \bar{\nu}_\tau$     & $1.47^{+0.43}_{-0.40}$  & $1.40^{+0.27}_{-0.26}$ &$1.31-1.48$    & $1.88\pm 0.20$\\
$B^- \to D^{*0} l^- \bar{\nu}_l$          & $4.87^{+1.60}_{-1.41}$  & $5.18^{+1.49}_{-1.28}$ &$5.35 \pm 0.16$& $5.69\pm 0.19$\\
\hline
\end{tabular}
\end{center} \end{table}

Since the large theoretical errors for the branching ratios  are largely canceled in the ratios
$R(D)$ and $R(D^*)$, one can define and calculate these two ratios rather than the branching ratios
with much better precision.
By using the ``pQCD + Lattice QCD" method we found the  following theoretical predictions:
\beq
R(D)& = & 0.337\pm 0.034(\omega_B)^{+0.017}_{-0.014}(f_B)\non
& = & 0.337^{+0.038}_{-0.037}, \non
R(D^*)& = &0.269\pm 0.018(\omega_B)^{+0.010}_{-0.009}(f_B)
\non
& = &0.269^{+0.021}_{-0.020},
\label{eq:rdt2}
 \eeq
where the remaining major theoretical errors come from the uncertainties
of $\omega_B=0.40\pm 0.04$ GeV and $f_B=0.21\pm 0.02$ GeV.
In Table \ref{tab:ratios}, we list the theoretical predictions for the $R(D^{(*)})$-ratios
obtained by using the pQCD approach \cite{csb59-125}, the ``pQCD+Lattice QCD" method (this work),
and the traditional HQET method \cite{prd85-094025}. In last column, we also list the
world averaged results as given by HFAG \cite{hfag-2015}.

\begin{table}[tb]
\begin{center}
\caption{ The theoretical predictions for the $R(D^{(*)})$-ratios.
In last three columns we list the measured values and their averages
\cite{prl109-101802,belle2015a,belle2015b}.}
\label{tab:ratios}\vspace{0.2cm}
\begin{tabular}{l lll  l  } \hline
Ratio \qquad  & pQCD \cite{csb59-125}\qquad & pQCD+Lattice \qquad& HQET \cite{prd85-094025} \qquad & HFAG
\cite{hfag-2015} \\ \hline
${\cal R}(D)$   & $0.430^{+0.021}_{-0.026}$&$0.337^{+0.038}_{-0.037}$& $0.296\pm 0.016$          & $0.391 \pm 0.041 \pm 0.028 $   \\
${\cal R}(D^*)$ & $0.301^{+0.013}_{-0.013}         $&$0.269^{+0.021}_{-0.020}$& $0.252\pm 0.003$ & $0.322 \pm 0.018\pm 0.012 $  \\
\hline
\end{tabular}
\end{center} \end{table}

\vspace{1cm}
\begin{figure}[tb]
\begin{center}
\vspace{0.3cm}
\centerline{\epsfxsize=12cm\epsffile{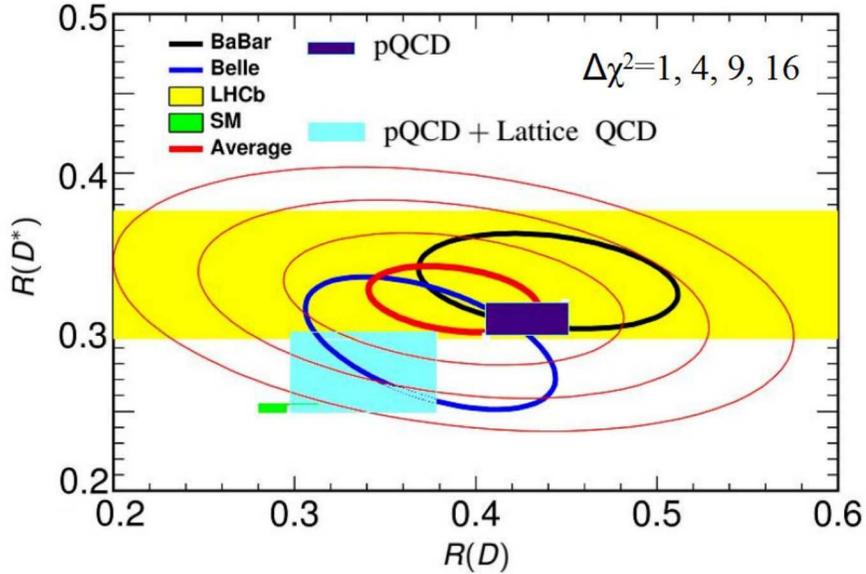}}
\caption{The 2-dimensional $R(D)-R(D^*)$ likelihood plane quoted directly
from Z.Ligeti's talk (see footnote 1) with the inclusion of the new theoretical predictions
obtained by using the pQCD approach (the purple box ) or the "pQCD + Lattice QCD" method (the
light-blue square) }
\label{fig:fig3}
\end{center}
\end{figure}

In Fig.~\ref{fig:fig3}, we draw the pQCD predictions for both ratios $R(D)$ and $R(D^*)$ directly on the
2-dimensional $R(D)-R(D^*)$ likelihood plane quoted directly from Z.~Ligeti's talk
\footnote{This is the figure appeared in Z.~Ligeti's talk given at the ICTP summer school on particle physics, June 15-26,
Trieste, Italy, and where the HQET predictions for $R(D^{(*)})$, the relevant experimental measurements
from BaBar, Belle and LHCb collaborations, and the world averages \cite{hfag-2015}
have been illustrated explicitly. },
where the purple square and the light-blue square added by us shows the theoretical predictions
with the errors obtained by using the pQCD approach \cite{csb59-125}  or the "pQCD + Lattice QCD"
method respectively.

From Table~\ref{tab:ratios} and Fig.~\ref{fig:fig3} we find the following points:
\begin{enumerate}
\item[(1)]
The errors of the pQCD predictions for the ratios $R(D^{(*)})$ become now around $ 10~\%$,
much smaller than those for the branching ratios.
But the current  $10~\%$ errors are about two times larger than the previous one ($\sim 5~\%$)
as given in Ref.~\cite{csb59-125}, since we assigned conservatively an overall error $\pm 0.02$ to the
Lattice QCD predictions for the values of the six form factors at the end-point $q^2=q^2_{\rm max}$
as given in Eq.~(\ref{eq:lqcds}).

\item[(2)]
From Fig.~3, one can see easily that our ``pQCD+Lattice QCD" predictions
for the ratios $R(D)$ and $R(D^*)$ do agree with the world averages within $2\sigma$ level,
This fact means that one can explain the $"R(D^{(*)})$-puzzle" without the aid
from the new physics beyond the SM.

\end{enumerate}

 For the $B\to D^{(*)} l \bar{\nu}_l$ decays, the pQCD predictions do depend on the distribution amplitudes of $D^{(*)}$ meson
as given in Eq.~(\ref{eq:phid}). In order to estimate the truncation effects of the Gegenbauer polynomials in
Eq.~(\ref{eq:phid}), we set the parameter $C_D=0$ and $C_{D^*}=0$ for the DAs in Eq.(\ref{eq:phid}),
and find the new pQCD predictions for the central values of the ratios $R(D)$ and $R(D^*)$:
\beq
R(D)=0.358, \quad R(D^*)=0.282.
\eeq
One can see that the pQCD predictions for $R(D^{(*)})$ become larger slightly than previous ones
in Eq.~(\ref{eq:rdt2}), $\sim 5~\%$ in magnitude and within one standard deviation. It is not hard to understand
such result: the changes of the form factors and then the branching ratios of the considered decays
induced by setting $C_{D^{(*)}}=0$ are largely canceled in the ratio $R(D^{(*)})$.

 \section{Summary} \label{sec:4}

In this paper, we studied the semileptonic decays $B \to D^{(*)} l^- \bar{\nu}_l$ in the framework of the
SM by employing the ``pQCD+Lattice QCD" method.
We made the extrapolation for the six relevant form factors $F_{(+,0)}(q^2)$, $V(q^2)$
and $A_{0,1,2}(q^2)$ by using the input values at sixteen points from the pQCD
factorization approach in the lower $q^2$ region ( $0\leq q^2 \leq m_\tau^2$),
and the ones at the end-point $q^2=q^2_{\rm max}$ from the Lattice QCD predictions as given in
Eq.~(\ref{eq:lqcds}).
We then calculated the branching ratios ${\cal B}(B \to D^{(*)} l^- \bar{\nu}_l)$ and
the ratio $R(D)$ and $R(D^*)$ by employing the new "pQCD+Lattice QCD" method.

From the numerical results and phenomenological analysis we found the following points:
\begin{enumerate}
\item[(1)]
By using the Lattice QCD predictions for the six form factors as the input at the end-point
$q^2=q^2_{\rm max}$, the reliability of the extrapolation of the form factors from the low to
the high $q^2$ region are improved effectively.

\item[(2)] The ``pQCD+Lattice QCD" predictions for the branching ratios
${\cal B}(B \to D^{(*)} l^-\bar{\nu}_l)$ agree well with the measured values within one standard deviation.

\item[(3)]
The  "pQCD+Lattice QCD" predictions for the ratios $R(D^{(*)})$ are the following
\beq
R(D)=0.337^{+0.038}_{-0.037}, \quad R(D^*)=0.269^{+0.021}_{-0.020}.
\label{eq:rdrds}
\eeq
They agree with the world  averages of those currently available experimental measurements
within $2\sigma$ deviation, in other words, one can explain the so-called "$R(D^{(*)})$-puzzle"
in the framework of the standard model.

\end{enumerate}


\begin{acknowledgments}
We wish to thank Hsiang-nan Li and Wen-Fei Wang for valuable discussions.
This work was supported by the National Natural Science
Foundation of China (11235005,11105115 and 11505148)  and
by the key project of Henan's Universities (15A140033).
\end{acknowledgments}







\end{document}